\documentclass{article}
\usepackage[utf8]{inputenc}
\usepackage{authblk}
\usepackage{graphicx}
\usepackage{amsmath}
\usepackage{bm}
\usepackage{blindtext}
\usepackage[a4paper, total={7.5in, 8.5in}]{geometry}
\usepackage{hyperref}
\hypersetup{
       colorlinks = true,
       citecolor = blue,
       linkcolor = blue,
       urlcolor =  blue,
}

\title{\textbf{Phase and Thermal Driven Transport across T-Shaped Double Quantum Dot Josephson Junction}}

\author{\bf Bhupendra Kumar \thanks{\bf bhupendra\_k@ph.iitr.ac.in}}
\author{Sachin Verma \thanks{sverma2@ph.iitr.ac.in}}
\author{Ajay \thanks{ajay@ph.iitr.ac.in}}
\affil{Department of Physics, Indian Institute of Technology, Roorkee, 247667, Uttarakhand, India}
\begin{document}

\maketitle
The phase and thermal driven transport properties of the T-shaped uncorrelated double quantum dot Josephson junction are analyzed by using Keldysh non-equilibrium Green's function equation of motion technique. In this setup, we have shown that the side-attached quantum dot provides an additional route for electron transmission which is affecting the transport properties by adjusting the interdot hopping between the main dot and the side dot. We began with investigating the impact of interdot hopping on Andreev bound states and Josephson supercurrent. When a small thermal bias is applied across the superconducting leads, the system exhibits a finite thermal response which is primarily due to the, thermally induced, quasi-particle current. The behavior of the Josephson supercurrent and the quasi-particle current flowing through the quantum dots is examined for various interdot hopping and thermal biasing. Finally, the system is considered in an open circuit configuration where the thermally driven quasi-particle current is compensated by the phase-driven Josephson supercurrent and the thermophase effect is observed. The effect of interdot hopping and the position of quantum dot energy level on the thermophase Seebeck coefficient is investigated.


\section{Introduction}
A quantum dot (QD)-based Josephson junction is made up of two Bardeen-Cooper-Schrieffer (BCS) superconducting leads separated by a quantum dot. A DC Josephson supercurrent can flow across the junction without applying potential difference, as the Josephson supercurrent largely depends on the phase difference between the superconductors \cite{Josephson1962, anderson1970josephson}. Quantum dots have discrete energy levels and can be controlled by tunning their gate voltage or by changing the size of quantum dot \cite{kouwenhoven2001few,Kastner1993}. Single-electron (quasi-particle) tunneling and cooper pair tunneling are responsible for charge transport in quantum dot-based Josephson junctions. Charge transport in these single quantum dot-based Josephson junctions have been studied extensively both theoretically \cite{rozhkov1999josephson,vecino2003josephson,choi2004kondo,lim2008andreev,zhu2001andreev,karrasch2008josephson,vodolazov2005superconducting,tanuma2007josephson,dhyani2009interplay} as well as experimentally \cite{van2006supercurrent,grove2007kondo,eichler2007even,ma2019andreev,pillet2013tunneling,lee2014spin,delagrange20160,szombati2016josephson}. Using quantum dots allow one to control the current flowing through Josephson junctions. Further, various authors have explored the charge transport properties of double quantum dot Josephson junctions. In such junctions, the double quantum dots are coupled with superconducting leads in series, parallel, and T-shaped geometry \cite{cheng2008josephson,chi2005current,zhu2002probing,lopez2007josephson,droste2012josephson,rajput2014tunable,lee2010josephson,saldana2018supercurrent}. References \cite{de2010hybrid,martin2011josephson,meden2019anderson} provides recent detailed reviews on the charge transport properties of single and double quantum dot based Josephson junctions. 

On the other hand, due to the limited temperature range, the thermal transport properties of the ordinary S-I-S Josephson junction and quantum dot-based junctions have not been widely explored. Despite this limitation, the thermal transport properties of Josephson junctions are recently attracting great attention \cite{guttman1997thermoelectric,giazotto2014proposal,martinez2015rectification,giazotto2015very,marchegiani2016self,marchegiani2020phase,bauer2021phase}.  Recently, very few studies have been conducted on the thermal transport properties of quantum dot-based Josephson junctions i.e when both the leads are superconducting \cite{kleeorin2016large,kamp2019phase}. Further, the thermoelectric transport properties of systems where the quantum dot is coupled between a normal metal and BCS superconductor (N-QD-S) \cite{Krawiec2008,hwang2015cross,verma2022non} and ferromagnet and BCS superconductor (F-QD-S) \cite{hwang2016hybrid,hwang2016large,hwang2017nonlinear,trocha2017spin}  have been studied recently. Further, the thermoelectric transport properties of multi-dot and multi-terminal systems with one superconducting lead are also gaining attention \cite{xu2016thermoelectric,yao2018enhancement,wysokinski2012thermoelectric,michalek2016local,hussein2019nonlocal}.

\begin{figure}
\centering
\includegraphics[width=0.8\textwidth]{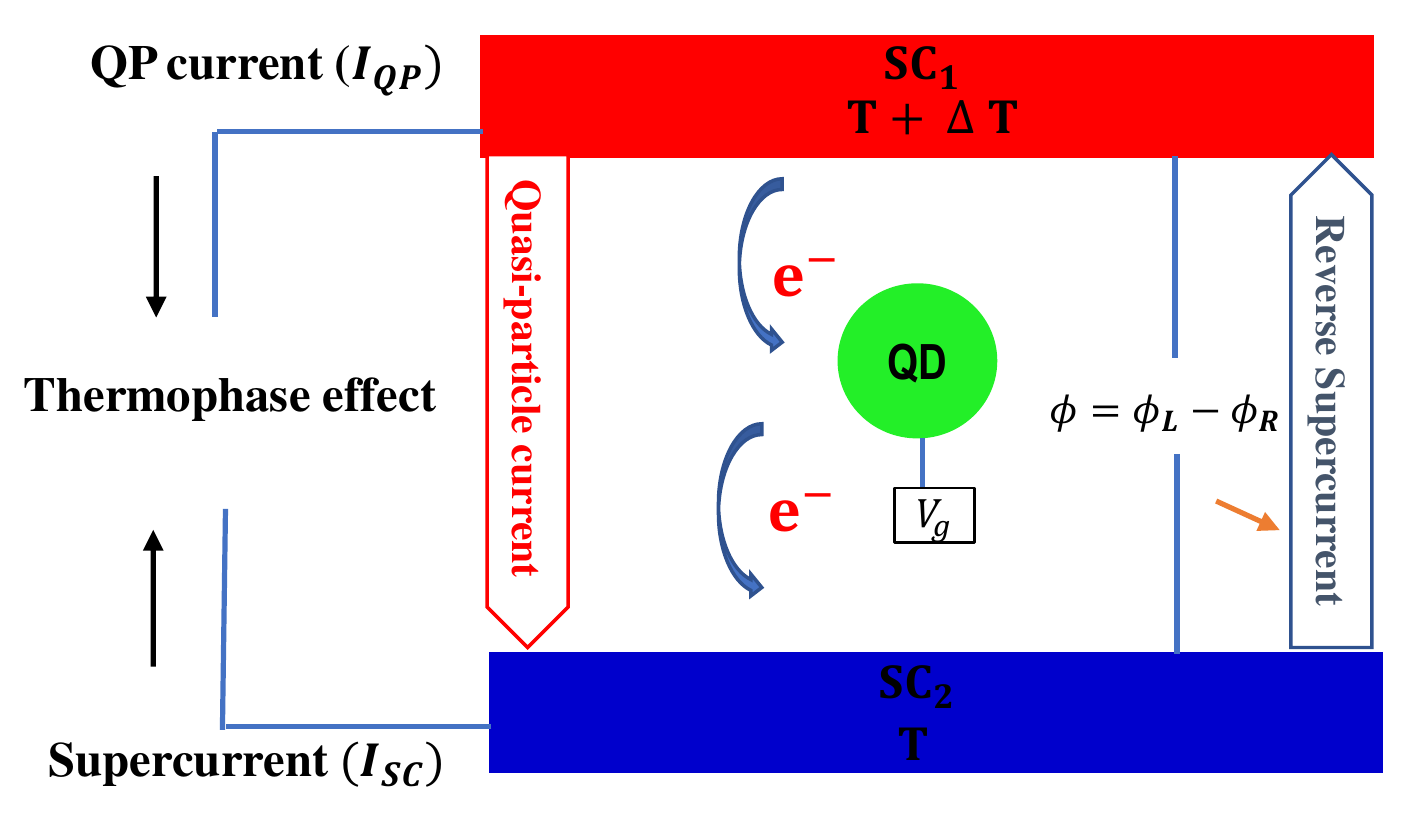}
\caption{\label{fig:fig_1}Schematic diagram showing the origin of thermophase Seebeck effect. The cancellation of quasi-particle current, induced by the temperature difference, by phase-driven reverse supercurrent is the origin of thermophase.}
\end{figure}

 Phase and thermal-driven transport properties of quantum dot-based Josephson junctions can be analyzed through a combination of three currents: quasi-particle current, interference current, and pair current \cite{guttman1997thermoelectric,kleeorin2016large,guttman1997phase}. A thermal gradient induces the quasi-particle current to flow across the junction. Quasi-particle is the only current that contributes to thermal transport in the S-QD-S system. The interference current, which is due to coupling between quasi-particle and condensate shows no contribution to thermal transport and will be ignored in the present study. The pair current or Josephson supercurrent flows across the junction in absence of voltage difference or temperature difference between the superconducting leads. This Josephson current depends on the phase difference between the superconducting leads. In reference, \cite{kleeorin2016large,kamp2019phase} author demonstrates that quantum dot-based Josephson junction shows a significant thermal response on applying the thermal biasing across the superconducting leads. By applying the thermal biasing across the superconducting leads, there appears to be a phase gradient across the superconductors. Therefore, a supercurrent will flow across the junction and it will counterbalance the thermally induced quasi-particle current. This is the open circuit configuration for S-QD-S system i.e. total current $I_{C}=0$. The cancellation of quasi-particle current by reverse supercurrent is the origin of concept of thermophase Seebeck effect in quantum dot-based Josephson junctions as shown by the schematic diagram in figure \ref{fig:fig_1}. 

\begin{figure}
\centering
\includegraphics[width=0.8\textwidth]{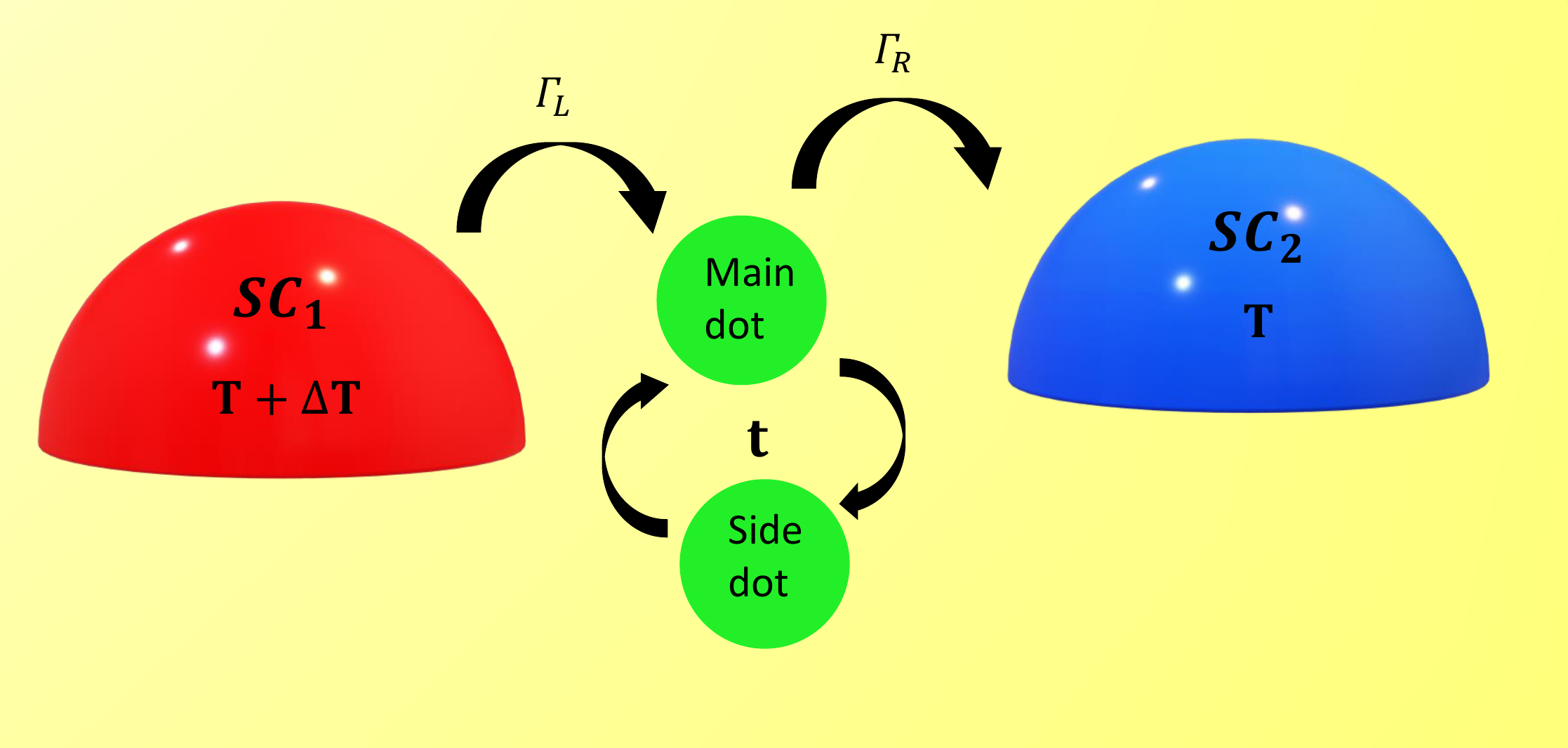}
\caption{\label{fig:fig_2}Schematic diagram for double quantum dot in T-shaped geometry coupled with superconducting leads. Main dot ($QD_{1}$) is directly coupled with superconducting leads while the side dot ($QD_{2}$) is only coupled with the main dot.}
\end{figure}

In the present work, we provide a study of the low-temperature phase and thermal-driven transport properties of a system where uncorrelated double quantum dots are coupled with two superconducting leads in T-shaped geometry (figure \ref{fig:fig_2}). In this configuration the main quantum dot ($QD_{1}$) is directly coupled with the leads and the side quantum dot ($QD_{2}$) is coupled with the main dot but not with the superconducting leads. To study the thermal transport properties of a T-shaped double quantum dot Josephson junction, we have employed Keldysh non-equilibrium Green’s equation of motion technique \cite{haug2008quantum,keldysh1965diagram}. First, we have studied the interdot hopping dependence of Andreev Bound States (ABS) and supercurrent. Next, total current (which is the combination of quasi-particle current and Josephson supercurrent) is calculated for different temperature differences $\Delta T$ and interdot hopping (t). Finally, the thermophase Seebeck coefficient (TPSC) for the T-shaped double quantum dot Josephson junction is analyzed. Since, both the leads are superconductors, so we have taken into account the temperature dependence of the superconducting gap ($\Delta_{\alpha}$) having a background temperature always less than the superconducting critical temperature $T_{c}$.
 
 This paper can be read in the following order: in the preceding section 2, we provide a detailed description of model Hamiltonian and theoretical formalism. Section 3, discusses numerical results. Lastly, section 4 concludes the present work.

\section{THEORETICAL FORMULATION}

To calculate the transport properties of the T-shaped double quantum dot Josephson junction, we use the generalized Anderson + BCS Hamiltonian in second quantization formalism.
\begin{equation}
    \hat{H}=\hat {H}_{leads}+\hat {H}_{QD}+\hat {H}_{tunnel}+\hat{H}_{interdot-hopping}
\end{equation}
where

\begin{equation*}
    \hat{H}_{leads} = \sum_{k\sigma,\alpha} \epsilon_{k\alpha}c^\dagger_{k\sigma,\alpha}c_{k\sigma,\alpha}-\left( \sum_{k\alpha}\Delta_{\alpha}c^\dagger_{k\uparrow,\alpha}c^\dagger_{-k\downarrow,\alpha}+h.c \right) 
\end{equation*}

\begin{equation*}
    \hat {H}_{QD}=\sum_{i=1}^{i=2}\sum_{\sigma}\epsilon_{d_{i\sigma}}d^\dagger_{i\sigma}d_{i\sigma}
\end{equation*}
\begin{equation*}
    \hat {H}_{tunnel}=\sum_{k\sigma,\alpha}V_{k,\alpha}c^\dagger_{k\sigma,\alpha}d_{1\sigma}+h.c
\end{equation*}

\begin{equation*}
    \hat{H}_{interdot-hopping} =\sum_{\sigma} t (d^\dagger_{1\sigma}d_{2\sigma}+h.c)
\end{equation*}
\\
 where h.c stands for Hermitian conjugate.
 
${\hat{H}}_{leads}$ is the Hamiltonian for left and right superconducting leads ($\alpha \in L,R$). The first term, describes the free electrons in the superconducting leads;  ${c^\dagger}_{k\sigma,\alpha}(c_{k\sigma,\alpha})$ is the creation (annihilation) operator of electron with spin $\sigma$ and wave vector $\vec{k}$ and energy $\epsilon_{k,\alpha}$. The second term in ${\hat{H}}_{leads}$ is the BCS term and gives the information about the interaction between Cooper pair with temperature dependent superconducting energy gap which is given by \cite{kamp2019phase}
 \begin{equation}
\Delta_{\alpha}(T_{\alpha})=\Delta_{0}\tanh \left\{ 1.74 \sqrt {\left( \frac{k_{B}T_{c}}{k_{B}T_{\alpha}}-1\right)}\right\}
\end{equation}
where $\Delta_{0}=\lvert \Delta_0 \rvert e^{i\phi_{\alpha}}$ is the superconducting gap at absolute zero temperature with $\phi_{\alpha}$ as the phase of superconducting leads, $T_{c}$ is superconducting critical temperature, $T_{\alpha}$ is the temperature of superconducting leads and $k_{B}$ is the Boltzman constant.
 
$\hat{H}_{QD}$ is the Hamiltonian for main dot ($QD_{1}$) and side dot ($QD_{2}$). $QD_{1}$ (i=1) and $QD_{2}$ (i=2) has energy $\epsilon_{d_{i\sigma}}$ with ${d^\dagger}_{i\sigma}(d_{i\sigma}$) as the fermionic creation operator (annihilation operator) of electrons with spin $\sigma$ and $n_{d_{i\sigma}} = d^\dagger_{i\sigma}d_{i\sigma}$ is the number operator. We have neglected the onsite Coulomb interaction $U_{i}$ for simplification.

$\hat{H}_{tunnel}$ is tunneling Hamiltonian between the energy level of the main dot and superconducting leads with interaction strength $V_{k_{1}},\alpha$. Further, we have consider the symmetric coupling strength of $QD_{1}$ to the left and right leads i.e. $V_{k_{1};L}$=$V_{k_{1};R}$. 
 
The last term $\hat{H}_{interdot-hopping}$ describes the interaction of electrons of two quantum dots via a hopping like term of strength t. Note that, there is no direct interaction between superconducting leads and $QD_{2}$.

 Bogoliubov transformation is used to diagonalize the BCS part of the Hamiltonian. For this, we introduce a new fermionic quasiparticle operator $\beta$ with coefficient
 $u_{k}$ and $v_{k}$ which satisfies the normalization condition  $\lvert u_{k} \rvert ^2$ + $\lvert v_{k} \rvert ^2$ = 1

 \begin{equation}
     c_{k\uparrow}=u_{k}^*\beta_{k\uparrow}+v_{k}\beta^\dagger_{-k\downarrow}
 \end{equation}
\begin{equation}
    c^\dagger_{-k\downarrow}=u_{k}\beta^\dagger_{-k\downarrow}-v^*_{k}\beta_{k\uparrow}
\end{equation}

By replacing the fermionic operator $ c_{k\uparrow}$ and $ c^\dagger_{-k\downarrow}$ with new quasi-particle operator, we get the effective Hamiltonian

\begin{equation}
\begin{aligned}
    \hat{H}= & \sum_{k\alpha}E_{k\alpha}(\beta^\dagger_{k\uparrow,\alpha}\beta_{k\uparrow,\alpha}+\beta^\dagger_{-k\downarrow,\alpha}\beta_{-k\downarrow,\alpha})
    \\ &
    + \sum_{k\alpha}(V_{k\alpha}u^*_{k}\beta^\dagger_{k\uparrow,\alpha}d_{1\uparrow} + V_{k\alpha}u^*_{k}\beta^\dagger_{-k\downarrow,\alpha}d_{1\downarrow}) 
    \\ &
    + \sum_{k\alpha}(V^*_{k\alpha}u_{k}d^\dagger_{1\uparrow}\beta_{k\uparrow,\alpha}+V^*_{k\alpha}u_{k}d^\dagger_{1\downarrow}\beta_{-k\downarrow,\alpha})
    \\ &
    +\sum_{k\alpha}V_{k\alpha}v_{k}(\beta_{-k\downarrow,\alpha}d_{1\uparrow}-\beta_{k\uparrow,\alpha}d_{1\downarrow}) 
    \\ &
    + \sum_{k\alpha}V^*_{k\alpha}v^*_{k}(d^\dagger_{1\uparrow}\beta^\dagger_{-k\downarrow,\alpha} - d^\dagger_{1\downarrow}\beta^\dagger_{k\uparrow,\alpha} ) 
   \\ &
   +   \epsilon _{d_{1}}(d^\dagger_{1\uparrow}d_{1\uparrow} + d^\dagger_{1\downarrow}d_{1\downarrow})
   \\ &
   + \epsilon _{d_{2}}(d^\dagger_{2\uparrow}d_{2\uparrow} + d^\dagger_{2\downarrow}d_{2\downarrow})
    \\ &
    + t (d^\dagger_{1\uparrow}d_{2\uparrow}+ d^\dagger_{1\downarrow}d_{2\downarrow} + d^\dagger_{2\uparrow}d_{1\uparrow} + d^\dagger_{2\downarrow}d_{1\downarrow})
\end{aligned}
\end{equation}

where $E_{k\alpha}=\sqrt{{\epsilon^2}_{k\alpha}+\lvert \Delta_{\alpha} \rvert^2}$ is excitation quasi-particle energy of the superconducting leads.The coefficients $u_{k}$ and $v_{k}$ can be expressed as

\begin{equation}
    \lvert u_{k} \rvert ^2= \frac{1}{2}\left(1+\frac{\epsilon_{k,\alpha}}{\sqrt{\epsilon^2_{k,\alpha}+\lvert\Delta_{\alpha}\rvert^2}}\right)
\end{equation}
\begin{equation}
    \lvert v_{k} \rvert ^2= \frac{1}{2}\left(1-\frac{\epsilon_{k,\alpha}}{\sqrt{\epsilon^2_{k,\alpha}+\lvert\Delta_{\alpha}\rvert^2}}\right)
\end{equation}

We have used Green's equation of motion method (EOM) to solve the above effective Hamiltonian (Eq. 5). To calculate the spectral and transport properties of T-shaped double quantum dot Josephson junction system, we need single-particle retarded Green's function of main quantum dot $QD_1$ which is defined in Zubarev notation \cite{haug2008quantum,keldysh1965diagram,zubarev1960double}
\begin{equation*}
    \langle\langle d_{\sigma}(t);d^\dagger_{\sigma}(0)\rangle\rangle=-i\theta(t)\langle[d_{\sigma}(t),d^\dagger_{\sigma}(0)]_{+}\rangle
\end{equation*}

The Fourier transform of the above retarded Green's function should satisfy the following equation of motion
\begin{equation}
    \omega\langle\langle d_{\sigma}\vert d^\dagger_{\sigma}\rangle\rangle_{\omega}=\langle\{d_{\sigma},d^\dagger_{\sigma}\}_{+}\rangle + \langle\langle[d_{\sigma}, H]_{-}\vert d^\dagger_{\sigma}\rangle\rangle_{\omega}
\end{equation}

In Nimbu space, Green's function of the main quantum dot can be represented by a $2\times2$ matrix:
\begin{equation}
\bm{G^r(\omega)}=
\begin{pmatrix}
    {G}^r_{11}(\omega) &  {G}^r_{12}(\omega)
     \\
     \\
      {G}^r_{21}(\omega) &  G^r_{22}(\omega)
\end{pmatrix}
=\begin{pmatrix}
    \langle\langle d_{1\uparrow}\vert d^\dagger_{1\uparrow}\rangle\rangle & \langle\langle d_{1\downarrow}\vert d_{\uparrow}\rangle\rangle
    \\
    \\
    \langle\langle d^\dagger_{1\downarrow}\vert d^\dagger_{1\uparrow}\rangle\rangle & \langle\langle d^\dagger_{1\downarrow}\vert d_{1\downarrow}\rangle\rangle
\end{pmatrix}
\end{equation}

By using Green's function EOM technique (Eq. 8), we obtain the following coupled equations for $QD_1$ 

\begin{equation}
\begin{aligned}
    (\omega-\epsilon_{d_{1}})\langle\langle d_{1\uparrow}\vert d^\dagger_{1\uparrow}\rangle\rangle  = & 1+ \sum_{k\alpha} V_{k\alpha} u^*_{k}\langle\langle \beta_{k\uparrow,\alpha}\vert d^\dagger_{1\uparrow}\rangle\rangle
    \\ & 
    + \sum_{k\alpha}V_{k\alpha}v_{k}\langle\langle \beta^\dagger_{-k\downarrow,\alpha}\vert d^\dagger_{1\uparrow}\rangle\rangle 
    \\ &
    + t \langle\langle d_{2\uparrow}\vert d^\dagger_{1\uparrow}\rangle\rangle 
\end{aligned}
 \end{equation}

\begin{equation}
\begin{aligned}
(\omega-E_{k\alpha})\langle\langle \beta_{k\uparrow,\alpha}\vert d^\dagger_{1\uparrow}\rangle\rangle = & \sum_{k\alpha} V^*_{k\alpha}u_{k}\langle\langle d_{1\uparrow}\vert d^\dagger_{1\uparrow}\rangle\rangle
\\ &
    + \sum_{k\alpha}V_{k\alpha}v_{k}\langle\langle d^\dagger_{1\downarrow}\vert d^\dagger_{1\uparrow}\rangle\rangle
\end{aligned}
\end{equation}

\begin{equation}
\begin{aligned}
    (\omega + E_{k\alpha})\langle\langle \beta^\dagger_{-k\downarrow,\alpha}\vert d^\dagger_{1\uparrow}\rangle\rangle = & \sum_{k\alpha} V^*_{k\alpha} v^*_{k} \langle\langle d_{1\uparrow}\vert d^\dagger_{1\uparrow}\rangle\rangle
  \\ &
    -\sum_{k\alpha} V_{k\alpha} u^*_{k} \langle\langle d^\dagger_{1\downarrow}\vert d^\dagger_{1\uparrow}\rangle\rangle
\end{aligned} 
\end{equation}

\begin{equation}
\begin{aligned}
(\omega + \epsilon_{d_{1}})\langle\langle d^\dagger_{1\downarrow}\vert d^\dagger_{1\uparrow}\rangle\rangle  = & \sum_{k\alpha} V^*_{k\alpha} v^*_{k}  \langle\langle \beta_{k\uparrow,\alpha}\vert d^\dagger_{1\uparrow}  \rangle\rangle
\\ &
   + \sum_{k\alpha} V^*_{k\alpha}u_{k} \langle\langle \beta^\dagger_{-k\downarrow,\alpha}\vert d^\dagger_{1\uparrow}\rangle\rangle 
\\ &
   - t \langle\langle d^\dagger_{2\downarrow}\vert d^\dagger_{1\uparrow}\rangle\rangle
\end{aligned}
\end{equation}

\begin{equation}
\begin{aligned}
  (\omega - \epsilon_{d_{2}})\langle\langle d_{2\uparrow}\vert d^\dagger_{1\uparrow}\rangle\rangle = & t\langle\langle d_{1\uparrow}\vert d^\dagger_{1\uparrow}\rangle\rangle 
\end{aligned}    
\end{equation}

\begin{equation}
\begin{aligned}
    (\omega - \epsilon_{d_{2}}) \langle\langle d^\dagger_{2\downarrow}\vert d^\dagger_{1\uparrow}\rangle\rangle = &  -t  \langle\langle d^\dagger_{1\downarrow}\vert d^\dagger_{1\uparrow}\rangle\rangle 
  \end{aligned}  
\end{equation}
 \\
After solving these closed set of coupled equations (Eq. 10-15) the expression for single particle retarded Green's function of the main dot with spin $\sigma=\uparrow$ can be written as:

\begin{equation}
    {G^r}_{11}(\omega)=\langle\langle d_{1\uparrow}\vert d^\dagger_{1\uparrow}\rangle\rangle=\frac{\omega+\epsilon_{d_{1}}- \frac{t^2}{\omega + \epsilon_{d_{2}}}-I_{1}} {(\omega+\epsilon_{d_{1}}- \frac{t^2}{\omega + \epsilon_{d_{2}}}-I_{1})(\omega-\epsilon_{d_{1}}- \frac{t^2}{\omega - \epsilon_{d_{2}}}-I_{2})-(I_{3})^2}
\end{equation}
\\
In above Green's function (Eq. 16) $I_{1}$, $I_{2}$ are the diagonal, and $I_{3}$ is the off-diagonal part of self-energy, which corresponds to the induced pairing, due to the coupling between the quantum dot and superconducting leads.
\\
The expressions for $I_{1}$, $I_{2}$ and $I_{3}$ are 
\begin{equation}
    I_{1} = \sum_{k\alpha}\lvert V_{k\alpha} \rvert ^2\left( {\frac{\lvert u_{k}\rvert ^2}{\omega+E_{k\alpha}}} + {\frac{ \lvert v_{k} \rvert ^2}{\omega-E_{k\alpha}}}\right)
\end{equation}
\begin{equation}
    I_{2} =  \sum_{k\alpha}\lvert V_{k\alpha} \rvert ^2 \left( {\frac{\lvert u_{k} \rvert ^2}{\omega-E_{k\alpha}}} + {\frac{\lvert v_{k} \rvert ^2}{\omega+E_{k\alpha}}} \right)
\end{equation}
\begin{equation}
I_{3} = \sum_{k\alpha} \lvert V_{k\alpha} \rvert ^2 u_{k}v^*_{k} \left( {\frac{1}{\omega-E_{k\alpha}}} - {\frac{1}{\omega+E_{k\alpha}}} \right)
\end{equation}
 Transforming the summation into integration and by defining the tunneling rate $\Gamma_{\alpha}=2\pi \rho_{0}\lvert V_{k\alpha} \rvert ^2$ where $\rho_{0}$ is the density of states in normal metallic state,  we obtained the following expressions for $I_{1}$, $I_{2}$ and $I_{3}$:
\begin{equation}
     I_{1}=I_{2}= -\sum_{\alpha \in L,R} \left (\frac{\Gamma_{\alpha}\omega}{\sqrt{\Delta_{\alpha} ^2 - \omega ^2}}\theta(\Delta-\lvert \omega\rvert)+ i(\frac{\Gamma_{\alpha}\omega}{\sqrt{\omega^2- \Delta_{\alpha}^2}}\theta(\lvert \omega\rvert-\Delta)
     \right )
\end{equation}
\begin{equation}
     I_{3}= -\sum_{\alpha \in L,R} \left (\frac{\Gamma_{\alpha}\Delta_{\alpha}}{\sqrt{ \Delta_{\alpha} ^2 - \omega ^2}}\theta(\Delta-\lvert\omega\rvert) + i(\frac{\Gamma_{\alpha}\Delta_{\alpha}}{\sqrt{\omega^2- \Delta_{\alpha}^2}}\theta(\lvert \omega\rvert-\Delta)   \right )
\end{equation}
\\
Similarly, the off-diagonal Green's function ${G^r}_{21}(\omega)$ can be calculated with the help of coupled equations (10-15).
\begin{equation}
\begin{aligned}
   {G^r}_{21}(\omega)=  \langle\langle d^\dagger_{1\downarrow}\vert d^\dagger_{1\uparrow}\rangle\rangle= \frac{I_{3}} {\big[({\omega + \epsilon_{d_{1\downarrow}}-\frac{t^2}{\omega + \epsilon_{d_{2\downarrow}}} + I_{1})({\omega - \epsilon_{d_{1\uparrow}}-\frac{t^2}{\omega - \epsilon_{d_{2\uparrow}}} + I_{2}) - (I_{3})^2\big]}}}
\end{aligned}
\end{equation}

Other Green's function can be obtained with the help of the following relations:
\begin{equation}
\begin{aligned}
{G}^r_{22}(\omega)= -{G}^r_{11}(-\omega)^*
\\
\\
{G}^r_{12}(\omega)= {G}^r_{21}(-\omega)^*
\end{aligned}
\end{equation}

By equating the denominator of single particle retarded Green's function ${G}^r_{11}$ (Eq. 16) equal to zero, the energies of Andreev Bound States (ABS) can be analyzed which is discussed in section III.

As discussed in section 1, for quantum dot-based Josephson junction, we can distribute the charge current in three parts \cite{guttman1997thermoelectric,kleeorin2016large}: 
\begin{equation}
\begin{aligned}
    I_C= & I_{QP} (\epsilon_{di},\phi,T,\Delta T) + I_{pair-QP} (\epsilon_{di},\phi,T,\Delta T) \cos^2 \frac{\phi}{2} 
    \\ & 
    + I_{SC} (\epsilon_{di},\phi,T,\Delta T) 
\end{aligned}
\end{equation} 
The first term is quasi-particle current and is responsible for thermal transport in this system. The second term has no contribution to thermal transport and can be neglected here. The third term in current is due to cooper pair tunneling and is responsible for the supercurrent in the system. Here, we present the full derivation of Josephson supercurrent and quasi-particle current.
For the current expression, we follow the formulation given in reference \cite{zhu2001andreev,haug2008quantum,meir1992landauer}.
The retarded Green's function $\bm{G^r}$ (Eq. 9) of main quantum dot can also be written as:
\begin{equation}
    \bm{G^r}=[\bm{g}^{r^{-1}} - \bm{\Sigma}^r]^{-1}
    =\frac{1}{A(\omega)} \begin{pmatrix}
  \bm{g}^r_{22}(\omega)-\bm{\Sigma}^r_{22} &  \bm{\Sigma}^r_{12}
\\
\\
 \bm{\Sigma}^r_{21} &  \bm{g}^r_{{11}}(\omega)-\bm{\Sigma}^r_{11}
\end{pmatrix}
\end{equation}

Where $\bm{G^r}$ is Green's function of the main quantum dot with leads and $\bm{g^r}$ is Green's function of the main quantum dot without leads. $\bm{\Sigma}^{r}=\bm{\Sigma}^r_{L}+\bm{\Sigma}^r_{R}$ is the retarded self-energy due to coupling between the main quantum dot and superconducting leads. $A(\omega)$ is the denominator of Green's function and is defined as 
\begin{equation}
    A(\omega)=det[\bm{g}^{r^{-1}}-\bm{\Sigma}^r]
\end{equation}

From Eq's. (9,16,22,23), $\bm{g}^r$ for T-shaped double quantum dot Josephson junction,  can be written as 
\begin{equation}
\bm{g}^{r^{-1}}=
    \begin{pmatrix}
        \omega-\epsilon_{d_{1}}-\frac{t^2}{\omega-\epsilon_{d_{1}}}+i0^{+} & 0
        \\
        \\
        0 &  \omega+\epsilon_{d_{1}}-\frac{t^2}{\omega+\epsilon_{d_{1}}}+i0^{+}
    \end{pmatrix}
\end{equation}
 The retarded self-energy ($\bm{{\Sigma}^r}_{\alpha}$) of $\alpha$ lead is defined as 
 \begin{equation}
    \bm{{\Sigma}^r}_{\alpha}(\omega)= -\frac{i}
    {2}\Gamma_{\alpha}\rho(\omega)  \begin{pmatrix}
  1 &  -\frac{\Delta}{\omega}e^{-i\phi_{\alpha}}
     \\
     \\
  -\frac{\Delta}{\omega}e^{i\phi_{\alpha}} &   1
\end{pmatrix}
\end{equation}
For simplicity, we assumed that both superconductors are identical and have a finite phase difference, $\Gamma_{\alpha}=\Gamma_L=\Gamma_R=\Gamma$, and $\Delta_{\alpha}=\Delta_{L}=\Delta_{R}=\Delta$
\begin{equation}
   \bm{\Sigma^r}=\bm{\Sigma^r}_{L} + \bm{\Sigma^r}_{R}=-i\Gamma\rho(\omega) \begin{pmatrix}
   1 & -\frac{\Delta}{\omega}cos\frac{\phi}{2}
   \\
   \\
   -\frac{\Delta}{\omega}cos\frac{\phi}{2} & 1
   \end{pmatrix}
\end{equation}

\begin{equation}
   \bm{\tilde{\Sigma}^r}=\bm{\Sigma^r}_{L} + \bm{\Sigma^r}_{R}=-i\Gamma\rho(\omega) \begin{pmatrix}
   0 & -\frac{\Delta}{\omega}(-i)sin\frac{\phi}{2}
   \\
   \\
   -\frac{\Delta}{\omega}(i)sin\frac{\phi}{2} & 0
   \end{pmatrix}
\end{equation}
where $\Gamma$ is the symmetric coupling strength between the main quantum dot and superconducting leads. $\rho(\omega)$ is the modified density of state of the superconductor and can be defined as

\begin{equation}
    \rho(\omega)= \begin{cases}
        \frac{\lvert\omega\rvert}{\sqrt{\omega^2-\Delta^2}} , \lvert\omega\rvert >\Delta 
        \\
        \\
         \frac{\lvert\omega\rvert}{i\sqrt{{\Delta^2}-\omega^2}} , \lvert\omega\rvert <\Delta 
\end{cases}
\end{equation}

By taking the current conservation condition $I_{L}+I_{R}=0$, the general formula of Josephson supercurrent for the superconductor-quantum dot systems can be written as \cite{zhu2001andreev}:
\begin{equation}
    I_{SC}=\frac{1}{2}(I_L-I_R)=\frac{e}{h}\int {d\omega}Re[\bm{G\tilde{\Sigma}}]^{<}_{11-22}
\end{equation}
Where $\bm{G}$ and $\bm{\Sigma}_{\alpha}$ are $2\times 2$ Fourier transformed Nambu matrices.

By using the Langreth relation we can write \cite{haug2008quantum};
\begin{equation}
    Re[\bm{G\tilde{\Sigma}}]^{<}_{11-22}=Re [\bm{{G^{<}}\tilde{\Sigma}^a}+\bm{G^r\tilde{\Sigma}^<}]_{11}-Re[\bm{{G^{<}}\tilde{\Sigma}^a}+\bm{G^r\tilde{\Sigma}^<}]_{22}
\end{equation}

with $(\bm{G^r})^\dagger=\bm{G^a}$, $\bm{(\tilde{\Sigma}^r)}^\dagger=\bm{\tilde{\Sigma}^r}$

 On applying the fluctuation-dissipation theorem, we can define:
 $\bm{G^{<}}=f(\omega)[\bm{G}^a-\bm{G}^r]$ and $\tilde{\bm{\Sigma}}^{<}=f(\omega)[\tilde{\bm{\Sigma}}^{a}-\tilde{\bm{\Sigma}}^{r}]$, where $f(\omega)=\frac{1}{e^{{\omega}\slash {k_{B}T}}+1}$ is Fermi distribution function and $\bm{G}^a$, $\bm{G}^{<}$ are advanced and lesser Green's function respectively.

Substituting the above relations in Eq. (33) and using Eq. (25) and Eq. (30)
\begin{equation}
    Re[\bm{G\tilde{\Sigma}^r}]^{<}_{11-22}=f(\omega)Re[(\bm{\tilde{\Sigma}}^r\bm{G}^r)^\dagger-\bm{G}^r\bm{\tilde{\Sigma}}^r]_{11}-f(\omega)Re[(\bm{\tilde{\Sigma}}^r\bm{G}^r)^\dagger-\bm{G}^r\bm{\tilde{\Sigma}}^r]_{22}
\end{equation}
\begin{equation}
    (\bm{\tilde{\Sigma}}^r\bm{G}^r)^{\dagger}_{11}= \frac{1}{A(\omega)} \begin{pmatrix}
   0 & -\frac{\Delta}{\omega}(-i)sin\frac{\phi}{2}
   \\
   \\
   -\frac{\Delta}{\omega}(i)sin\frac{\phi}{2} & 0
   \end{pmatrix} 
  \begin{pmatrix}
  \bm{g^r}_{22}(\omega)-\bm{\Sigma}^r_{22} &  \bm{\Sigma}^r_{12}
     \\
     \\
 \bm{\Sigma}^r_{21} &   \bm{g}^r_{11}(\omega)-\bm{\Sigma}^r_{11}
\end{pmatrix}
\end{equation}
Similarly, we obtain the $(\bm{G^{r}}\bm{\tilde{\Sigma}^r})_{11}$ and the $Re[(\bm{\tilde{\Sigma}}^r\bm{G}^r)^\dagger-\bm{G}^r\bm{\tilde{\Sigma}}^r]_{22}$ in Eq. (34). Finally it can be written as:
\begin{equation}
    Re[\bm{G\tilde{\Sigma}}]^{<}_{11-22}= \frac{\Gamma^2\Delta^2}{\omega^2-\Delta^2}f(\omega)sin{\phi} Im\left [\frac{-1}{A(\omega)} \right ]
\end{equation}
By substituting Eq. (36) in Eq. (32) and taking both spin up and spin down into consideration, the final expression for Josephson supercurrent can be written as
\begin{equation}
I_{SC}=\frac{2e}{h}\int{d{\omega}}\frac{\Gamma^2\Delta^2}{\omega^2-\Delta^2}f(\omega)sin{\phi} Im \left [ \frac{-1}{A(\omega)} \right ]
\end{equation}

Similiar to Josephson supercurrent, we can also derive the expression of quasi-particle current for superconductor-quantum dot systems by following the formalism given in reference \cite{haug2008quantum,kang1998transport,meir1992landauer}.

\begin{equation}
    I_{\alpha}=\frac{4ie}{h}\int {d\omega}\Gamma_{\alpha}\frac{\lvert\omega\rvert}{\sqrt{\omega^2-\Delta^2_{\alpha}}} \{f_{\alpha}(\omega)[G^r_{11}(\omega)-G^a_{11}(\omega)]+G^{<}_{11}(\omega)\}
\end{equation}
By current conservation condition i.e. $I_L+I_R=0$, the symmetrize form of current can be written as 
\begin{equation*}
     I_{QP}=I_L=-I_R=\frac{1}{2}(I_L-I_R)
\end{equation*}
    
\begin{equation}
\begin{split}
     I_{QP}=\frac{2ie}{h} \int{d\omega}\{[\Gamma_{L}\frac{\lvert\omega\rvert}{\sqrt{\omega^2-\Delta^2_{L}}} f_{L}(\omega)- \Gamma_{R}\frac{\lvert\omega\rvert}{\sqrt{\omega^2-\Delta^2_{R}}} f_{R}(\omega)]
     [G^r_{11}(\omega)-G^a_{11}(\omega)]
     \\ 
     +[\Gamma_L(\omega)-\Gamma_R(\omega)]G^{<}_{11}(\omega)\}
\end{split}
\end{equation}
For the symmetric case ($\Gamma_{\alpha}=\Gamma_{L}=\Gamma_{R}=\Gamma$ and $\Delta_{\alpha}=\Delta_{L}=\Delta_{R}=\Delta$) the above equation becomes
\begin{equation}
\begin{aligned}
  I_{QP}= \frac{2ie}{h}\int d{\omega}{\Gamma}\frac{\lvert\omega\rvert}{\sqrt{\omega^2-\Delta^2}} \{f_{L}(\omega)-f_{R}(\omega)\}[G^r_{11}(\omega)-G^a_{11}(\omega)]
\end{aligned}
\end{equation}
where 
   $f_{L}(\omega) =\frac{1}{e^{{\omega}\slash {k_{B}(T+\Delta T)}}+1}$, $f_{R}(\omega)=\frac{1}{e^{{\omega}\slash {k_{B}T}}+1}$ are the Fermi-distribution function of left and right superconducting leads and $[G^r_{11}(\omega)-G^a_{11}(\omega)]=Im[-G^r_{11}(\omega)]$.

The expression of quasi-particle current for linear response regime (thermal gradient between superconducting leads will be small, $\Delta T\rightarrow 0$) can be simplified by using power series expansion of the Fermi-distribution function of left and right superconducting leads:
\begin{equation}
I_{QP}=\frac{2e}{h} \int{d\omega}\frac{df(\omega)}{dT}\Gamma Re({\rho(\omega)})Im[-G^r_{{11}}(\omega)]\Delta T
\end{equation}

In section 1, we have already discussed the origin of the thermophase effect. Within the linear response regime, the thermophase Seebeck coefficient $S_{\phi}$ is defined analogous to the thermovoltage Seebeck coefficient and can be simplified using Eq. (24);
\begin{equation}
S_{\phi}=-\left(\frac{\Delta \phi}{\Delta T}\right)_{I=0} = \frac{dI_{QP}\slash{d\Delta T}}{I_{SC}}
\end{equation}

\section{RESULTS and DISCUSSION}

In this section, we present the numerical results and discussion for the T-shaped double quantum dot Josephson junction. Transport properties are discussed for uncorrelated quantum dots. The superconducting gap at absolute zero temperature ($\Delta_{0}$) is considered as the energy unit, where $\Delta_{0}$ is in meV.

\begin{figure} [ht]
\centering
\includegraphics[width=0.8\textwidth]{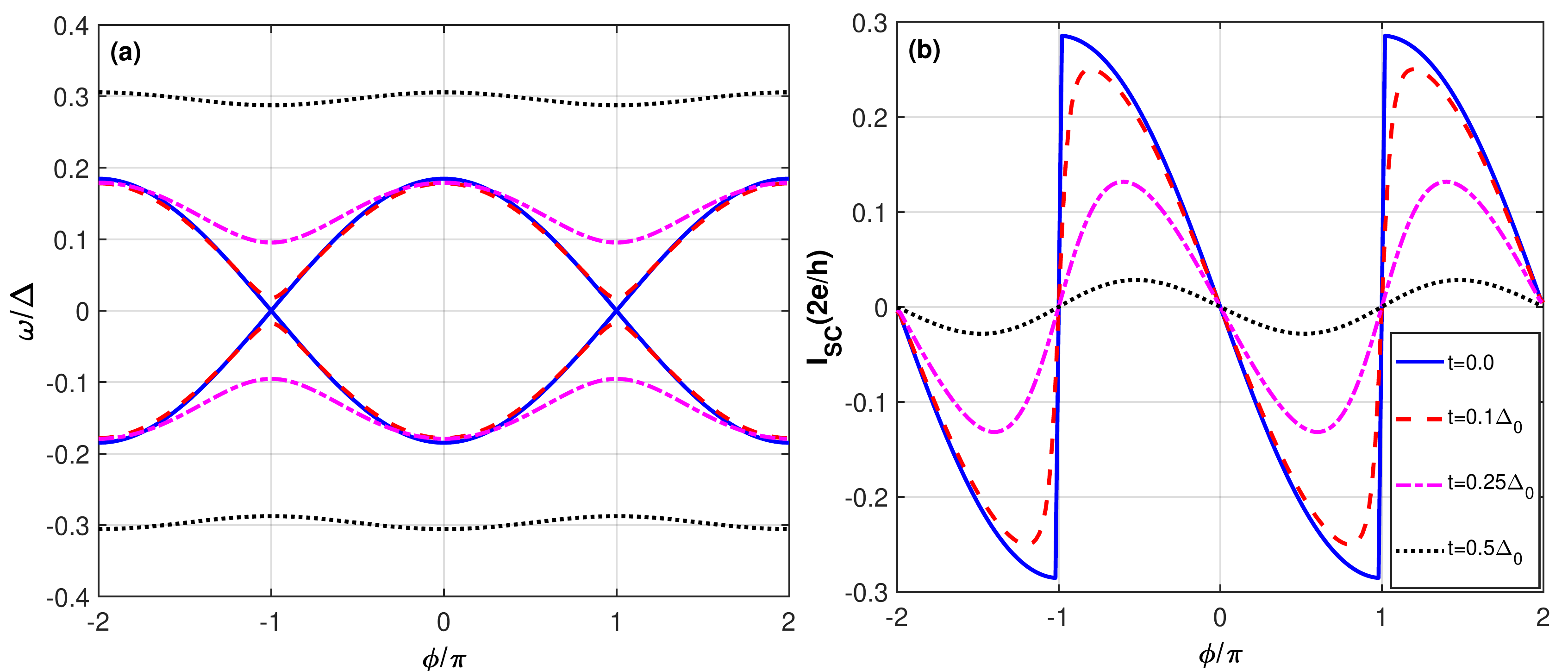}
\caption{\label{fig:fig_3}(a) Energy of Andreev bound states and (b) Josephson current as a function of superconducting phase difference $(\phi)$ for different values of interdot hopping (t) at absolute zero temperature. The other parameters are $\Gamma =0.1 \Delta_{0}$, $\epsilon_{d_{1}}=0$, $\epsilon_{d_{2}} = 0.5 \Delta_{0}$.}
\end{figure}
In figure \ref{fig:fig_3}, we plot the energy of Andreev Bound states (ABS) and Josephson current as a function of superconducting phase difference ($\phi$) for different values of interdot hopping (t). First, when $ QD_{2} $ is decoupled from $QD_{1}$ i.e. t=0, the system shows the properties of the usual S-QD-S Josephson junction. In this case, the supercurrent is discontinuous at $\phi=\pm \pi$ and upper and lower ABS crosses the Fermi energy ($\omega=0$). Thus for t=0, the system acts as a perfect transmitting channel. When $ QD_{2} $ is coupled with $ QD_{1} $ then supercurrent shows sinusoidal behaviour and a finite gap is generated between lower and upper ABS at $\phi=\pm \pi$. Further increasing the value of interdot hopping, this supercurrent is suppressed. This suppression of supercurrent is because coupling $QD_{1}$ with the $QD_{2}$, the electrons tends to tunnel into $QD_{2}$. This causes interference destruction between two transport channels and as a result the supercurrent decreases. Thus for $t > 0$, the system does not acts as a perfect transmitting channel. The suppression of supercurrent can also be explained in terms of the splitting of QDs energy level due to interdot hopping. When $t\neq 0$, the equivalent level of QDs splits into two levels i.e. $\bar{\epsilon_{d_i}} = \epsilon_{d_{i}} \pm t$. The equivalent energy level $\bar{\epsilon_{d_{i}}}$ moves far away from the Fermi level with increasing interdot hopping and supercurrent decreases.
\begin{figure} [!htb]
\centering
\includegraphics[width=0.8\textwidth]{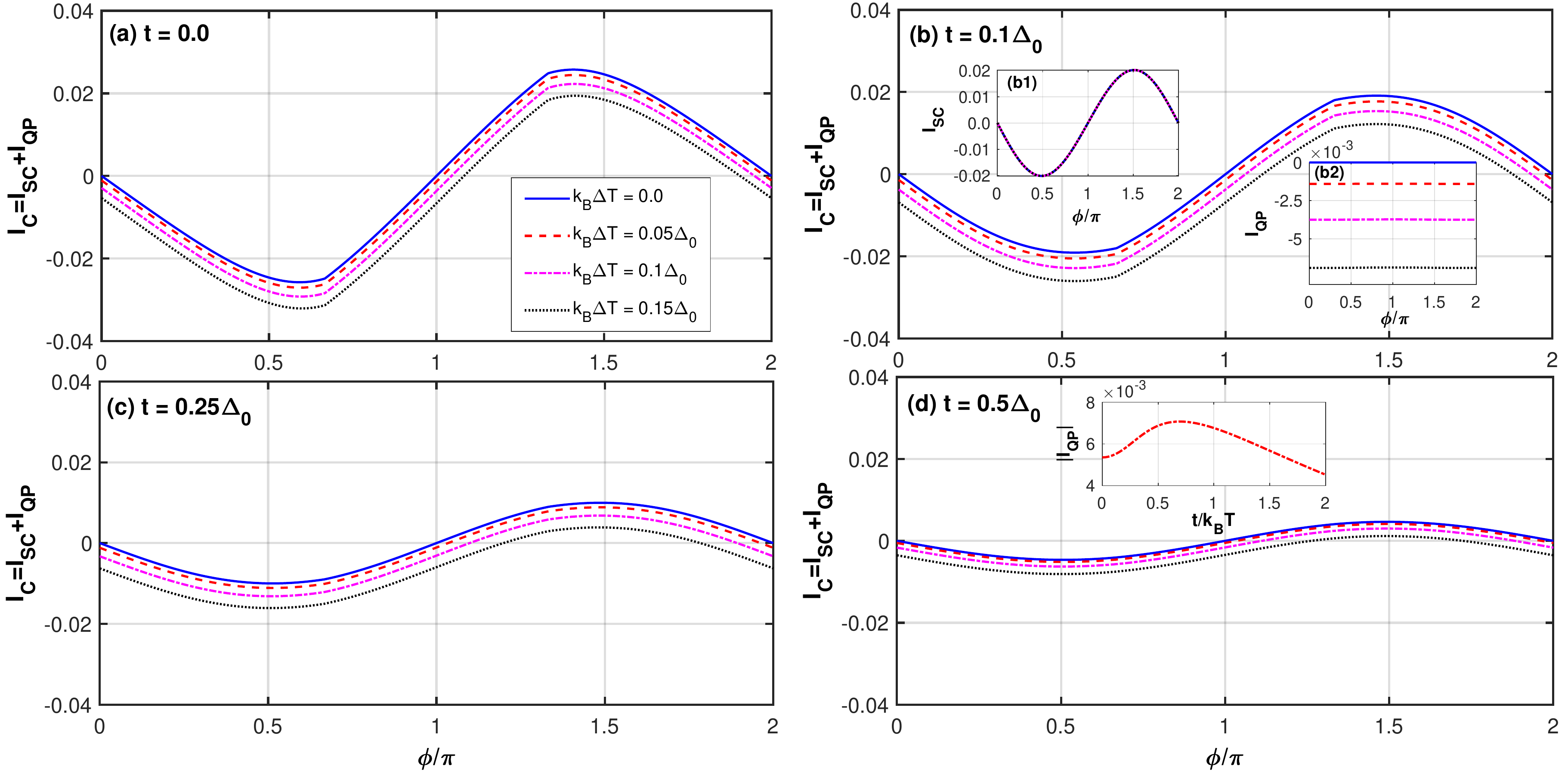}
\caption{\label{fig:fig_4}Total current (supercurrent + quasi-particle current) as a function of superconducting phase difference $(\phi)$ for different interdot hopping (t) and thermal biasing $\Delta T$. Insets ($b_{1}$) and ($b_{2}$) show the separate behaviour of Josephson current and quasi-particle current with $\phi$ and $\Delta T$. Inset in figure (d) shows the variation of quasi-particle current as a function of interdot hopping. The other parameters are $\Gamma=0.1 \Delta_{0}$, $k_{B}T=0.2 \Delta_{0}$, $\epsilon_{d_{1}}=\epsilon_{d_{2}}=-1.0 \Delta_{0}$.}
\end{figure}

In figure \ref{fig:fig_4} [(a)-(d)], we plot the total current as a function of superconducting phase difference ($\phi$) for several values of thermal biasing ($\Delta T$) and interdot hopping. First, by decoupling $QD_{2}$ form $QD_{1}$ (Fig 4 a) the results of S-QD-S system are reproduced \cite{kleeorin2016large}. When both $QD_{1}$ and $QD_{2}$ are coupled, then there is suppression in the magnitude of total current with increasing interdpt hopping, which is discussed in the previous paragraph. In insets 4(b), the individual behavior of Josephson current and quasi-particle current are shown. It is observed that Josephson current is almost independent of thermal biasing ($\Delta T$) and largely depends on superconducting phase difference ($\phi$). On other hand, the quasi-particle current totally depends on the thermal biasing ($\Delta T$). It is important to note that in total current, the sinusoidal nature is due to Josephson current, and the shift in magnitude is due to quasi-particle current. With an increase in interdot hopping, the amplitude of the supercurrent and total current vanishes. In inset 4(d), the behaviour of quasi-particle current is plotted as a function of interdot hopping. It is observed that quasi-particle current first increases with interdot hopping and then attain a maximum value for $k_{B}T \sim {t}$, and then decreases with a further increase in the value of interdot hopping.
 
\begin{figure} [!htb]
\centering
\includegraphics[width=0.8\textwidth]{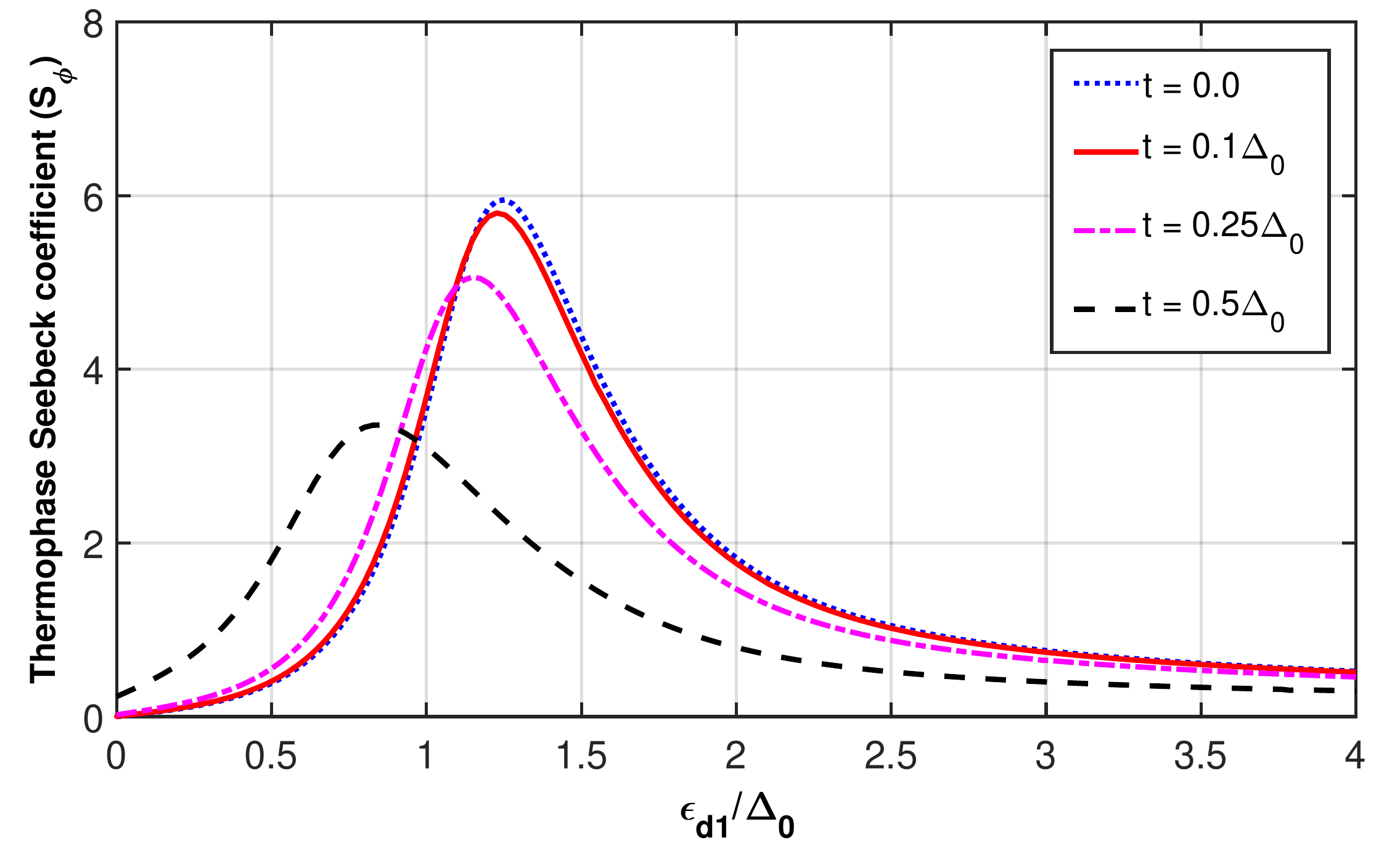}
\caption{\label{fig:fig_5}The variation of thermophase Seebeck coefficient ($ S_{\phi}$) with $QD_{1}$ energy level $\epsilon_{d_{1}}$ for different interdot hopping (t). The other parameters are $\Gamma=0.1\Delta_{0}$, $k_{B}T=0.2\Delta_{0}$, $\epsilon_{d_{2}}=0.5\Delta_{0}$.}
\end{figure}
 
\begin{figure} [t]
\centering
\includegraphics[width=0.8\textwidth]{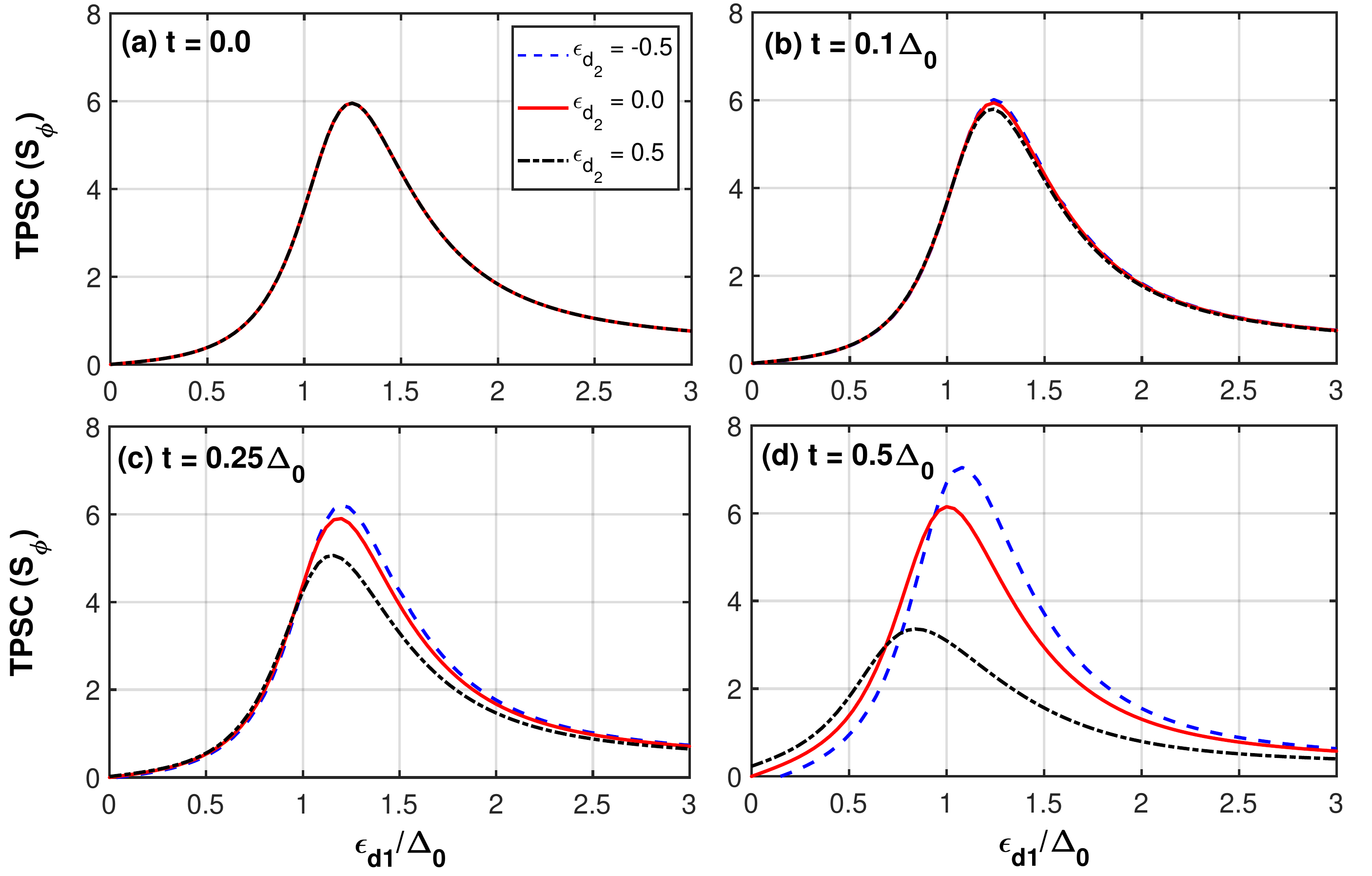}
\caption{\label{fig:fig_6}The variation of thermophase Seebeck coefficient ($ S_{\phi}$) with $QD_{1}$ energy level $\epsilon_{d_{1}}$ for different values of $QD_{2}$ energy level $\epsilon_{d_{2}}$ and $t$. The other parameters are $\Gamma=0.1\Delta_{0}$, $k_{B}T=0.2\Delta_{0}$.}
\end{figure}

As discussed previously, the origin of the thermophase effect is due to the vanishing total current in open circuit configuration i.e. thermal-driven quasi-particle current is compensated by the phase-driven Josephson supercurrent flowing in the reverse direction. In figure \ref{fig:fig_5} and \ref{fig:fig_6}, we have analyzed thermophase Seebeck coefficient $(S_{\phi})$ of S-TDQD-S in linear response regime for uncorrelated quantum dots.

In figure \ref{fig:fig_5}, we have plotted the thermophase Seebeck coefficient (TPSC) as a function of $QD_{1}$ energy level for different interdot hopping. It is observed that TPSC ($S_{\phi}$) peaks are highest for t=0 i.e., when $QD_{2}$ is decoupled from $QD_{1}$ \cite{kleeorin2016large}. When $QD_{2}$ is coupled with $QD_{1}$, TPSC peaks start decreasing with increasing interdot hopping (t). To achieve a high thermophase peak Josephson current should compensate the quasi-particle current totally.  As we have shown that Josephson supercurrent decreases with increasing interdot hopping, therefore it compensates less quasi-particle current. Thus, TPSC peaks decrease and produce a shift in peaks with increasing interdot hopping.

The magnitude of TPSC not only depends on interdot hopping but also depends on the position of $QD_{2}$ energy levels whether it lies above or below the Fermi level. In figure \ref{fig:fig_6}, we plot the thermophase Seebeck coefficient ($S_{\phi}$) as a function of $QD_{1}$ energy level for different values of $QD_{2}$ energy level. When $QD_{2}$ is decoupled from $QD_{1}$, then system reproduces the results of S-QD-S for TPSC ($S_{\phi}$). For finite interdot hopping, the magnitude of TPSC peaks are enhanced when $QD_{2}$ energy level lies below the Fermi level. The enhancement of TPSC peaks can be explained as follows: when $QD_{2}$ energy level lies below the Fermi level, the equivalent or effective level lies close to the Fermi level which supports the resonant cooper pair tunneling. Therefore thermally induced quasi-particle is compensated by Josephson supercurrent completely i.e. large magnitude of TPSC peaks. These results for different values of $QD_{2}$ energy levels can be directly compared with the TPSC plots as discussed in figure (5).

\section{CONCLUSION}

We have addressed the phase-driven and thermal-driven transport properties through a T-shaped double quantum dot Josephson junction. For uncorrelated quantum dots, the impact of interdot hopping on Andreev bound states (ABS) and Josephson supercurrent are investigated. For a finite value of interdot hopping, Josephson supercurrent exhibits sinusoidal nature while ABS shows a finite gap around the Fermi level. The magnitude of Josephson supercurrent decreases with increasing interdot hopping because the electrons have a tendency to tunnel into the side dot with increasing interdot hopping, which results in interference destruction between two transport channels. Further, this system exhibits a finite thermal response when a small thermal biasing ($\Delta T$) is applied across the superconducting leads. The quasi-particle current flows across the junction due to thermal biasing, while the Josephson current is almost insensitive to thermal biasing. With increasing thermal biasing, the quasi-particle current produces a finite shift in the magnitude of the total current. Also, the magnitude of the total current decreases with increasing interdot hopping because the equivalent level of quantum dots moves further away from the Fermi level and thus cooper pair tunneling is suppressed.

Finally, we investigate the influence of interdot hopping and quantum dot energy levels on the thermophase Seebeck coefficient (TPSC). The magnitude of TPSC ($S_{\phi}$) decreases with increasing interdot hopping when the energy levels of the side dot ($QD_{2}$) lie above the Fermi level and increase when energy levels of the side dot ($QD_{2}$) lie below the Fermi level. In the later case, when energy levels lie below the Fermi level, the equivalent level ($\epsilon_{d_{i}} \pm t$) of the quantum dot moves towards the Fermi level. Thus, the cooper pair tunneling increases with interdot hopping, and quasi-particle current is completely compensated by the Josephson current and the magnitude of TPSC peaks is enhanced.

We believe that the results presented in this study can be tested experimentally with the advancement in nano-fabrication techniques. In this paper, we consider the uncorrelated quantum dots, and the effect of the Coulomb correlation will be explored in future work. The present study can also be extended to investigate the thermal transport properties in systems where double quantum dots are coupled with superconducting leads in series, and parallel geometry and also for multi-terminal configurations. The concept of thermophase effect in quantum dot-based Josephson junction can be useful for future low-temperature thermal applications \cite{giazotto2006opportunities,quan2006maxwell,martinez2014coherent,fornieri2016nanoscale} and need further investigation.
\\
\\
\textbf{ Acknowledgments }

The authors acknowledge the financial support from the research project DST-SER-1644-PHY 2021-22. Bhupendra Kumar also acknowledges the support from the Ministry of Education (MoE), India, in the form of a Ph.D. fellowship.

\bibliographystyle{unsrt}  
\bibliography{MyBib}

\end{document}